\documentclass[twocolumn,superscriptaddress,10pt,pra,showpacs]{revtex4}
\usepackage{amsmath,graphicx,amsfonts,bm,amssymb}
\usepackage{times}
\begin{document}

\title{Implementing topological quantum manipulation with superconducting circuits}

\author{Zheng-Yuan Xue}

\affiliation{Department of Physics and Center of Theoretical and
Computational Physics,\\ The University of Hong Kong, Pokfulam Road,
Hong Kong, China}

\author{Shi-Liang Zhu}

\affiliation{Laboratory of Quantum Information Technology, ICMP and
SPTE, South China Normal University, Guangzhou 510006, China}

\affiliation{Department of Physics and Center of Theoretical and
Computational Physics,\\ The University of Hong Kong, Pokfulam Road,
Hong Kong, China}

\author{J. Q. You}
\affiliation{Department of Physics and Surface Physics Laboratory
(National Key Laboratory),\\ Fudan University, Shanghai 200433,
China}

\author{Z. D. Wang}

\affiliation{Department of Physics and Center of Theoretical and
Computational Physics,\\ The University of Hong Kong, Pokfulam Road,
Hong Kong, China}

\date{\today}

\begin{abstract}
A two-component fermion model with conventional two-body
interactions was recently shown to have anyonic excitations. We here
propose a scheme to physically implement this model by transforming
each chain of two two-component fermions to the two capacitively
coupled chains of superconducting devices. In particular, we
elaborate how to achieve the wanted operations to create and
manipulate the topological quantum states, providing an
experimentally feasible scenario to access the topological memory
and to build the anyonic interferometry.
\end{abstract}


\maketitle


Topological ordered states emerge as a new kind of states of quantum
matter beyond the description of conventional Landau's theory
\cite{wenbook}, whose excitations are anyons satisfying fractional
statistics. A paradigmatic system for the existence of anyons is a
kind of so-called fractional quantum Hall states \cite{fqhs}.
Alternatively, artificial spin lattice models are also promising for
observing these exotic excitations \cite{wenbook,kitaev97,kitaev06}.
Kitaev models \cite{kitaev97,kitaev06} are most famous for
demonstrating anyonic interferometry  and braiding operations for
topological quantum computation.

Since anyons have not been directly observed experimentally, a focus
at present  is to experimentally demonstrate the topological nature
of these states. Abelian anyons maybe relatively easy to achieve and
to manipulate in comparison with the nonabelian ones, thus it is of
current interest to explore them both theoretically and
experimentally. Kitaev constructed an artificial spin model
\cite{kitaev97}, i.e., the toric code model, which supports the
abelian anyon. But the wanted four-body interactions are notoriously
hard to generate experimentally in a controllable fashion.
Alternatively, it was proposed \cite{han} to generate dynamically
the ground state and the excitations of the model Hamiltonian,
instead of direct ground-state cooling, to simulate the anyonic
interferometry.  On the other hand, implementation of another
Kitaev's honeycomb model \cite{kitaev06} was also suggested in the
context of ultracold atoms \cite{duan}, polar molecules
\cite{zoller}, and superconducting circuits \cite{youtopo}. The
honeycomb model \cite{kitaev06} is an anisotropic spin model with
three types of nearest-neighbor two-body interactions, which support
both abelian and nonabelian anyons. It was shown \cite{kitaev06}
that the toric code model can be obtained from the limiting case of
the honeycomb model. Using this map, preliminary operations for
topological quantum memory and computation were also addressed
\cite{zhang,jiang,cirac,vidal}. Nevertheless, in this case, anyons
are created by the fourth-order perturbation treatment, which would
blur the extracted anyonic information \cite{vidal}. In addition,
this map is good but not exact, so that one may get both anyonic and
fermionic excitations. Therefore, new methods for implementing the
model and manipulating the relevant topological states are still
desirably awaited.

Recently, a two-component fermion model \cite{yu} with conventional
two-body interactions was shown to have anyonic excitations, which
obey the same fusion rules as those of the toric code model and are
mutual semions. This model is promising because it provides an
example for abelian anyons, which can be directly implemented. In
this Rapid Communication, we propose to physically implement this
model with appropriately designed superconducting circuits. In
particular, we elaborate how to achieve the wanted operations that
create and manipulate the topological states as well as anyons with
the cavity-assisted interactions using an external magnetic drive,
providing an experimentally feasible scenario to access the
topological memory and to build the anyonic interferometry.

\begin{figure}[tbp]
\centering
\includegraphics[width=7.5cm]{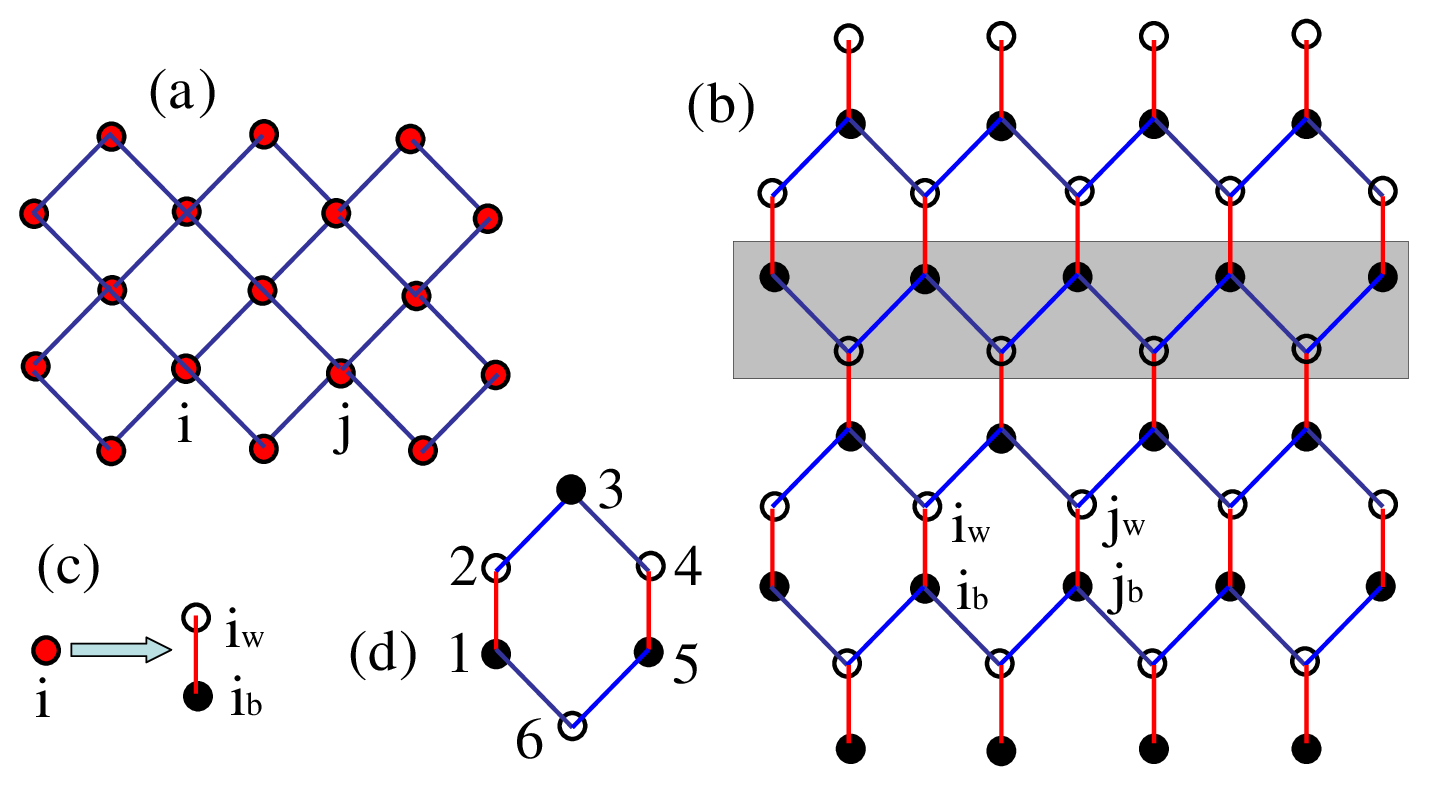}
\caption{(Color online) A map from a square lattice  to a honeycomb
lattice by extending each square lattice site to be a vertical link.
(a) The square lattice, where $\langle i, j\rangle$ denotes the
nearest neighbors along the  horizontal diagonal direction. (b) The
honeycomb lattice, where $i_{b,w}$ label the black and white
sublattices. The indicated zig-zag chain is one of the lines for the
Jordan-Wigner transformation. (c) Each square lattice site being
extended as a vertical link of Majorana fermions with  $i_w$ ($i_b$)
labels the white (black) sublattice and $i_w$ on the top of $i_b$.
(d) Site-labels within a plaquette.} \label{map1}
\end{figure}

The Hamiltonian for the two-component fermion model in a
two-dimensional square lattice, as shown in Fig. \ref{map1}(a), is
\cite{yu}
\begin{eqnarray} \label{h}
H_{f1}&=&-J_q\sum_{\langle
i, j\rangle} (2n_{\uparrow,i}-1)(2n_{\uparrow,j}-1)\nonumber\\
&& -J_p\sum_{\langle i, j\rangle}
(2n_{\downarrow,i}-1)(2n_{\downarrow,j}-1)\nonumber\\
&&+U\sum_{i}(2n_{\uparrow,i}-1)(2n_{\downarrow,i}-1),
\end{eqnarray}
where $n_{s,i}=c^\dag_{s,i}c_{s,i}$, $c_{s,i}$ are annihilation
operators of spin-$s$ fermions, and $\langle i, j\rangle$ mean the
nearest neighbors along the horizontal diagonals of squares. The
ground states of this Hamiltonian are highly degenerated, i.e.,
every individual chain is ferromagnetic. As shown in Ref. \cite{yu},
the low-lying excitations  are deconfined mutual semions under the
open boundary condition.

To show the nature of the low-lying excitations, we map the square
lattice to the honeycomb lattice, as shown in Fig. \ref{map1}(b), by
extending each lattice site in the square lattice to be a vertical
link of Majorana fermions [see Fig. \ref{map1}(c)] defined by
$\psi_{i_b}=-i(c_{\uparrow,i}-c_{\uparrow,i}^\dag)$,
$\psi_{i_w}=c_{\uparrow,i}+c_{\uparrow,i}^\dag$;
$\chi_{i_b}=-i(c_{\downarrow,i}-c_{\downarrow,i}^\dag)$, and
$\chi_{i_w}=c_{\downarrow,i}+c_{\downarrow,i}^\dag$. These "real"
fermion operators obey $\psi_{i_s}^2=\chi_{i_s}^2=1$. Otherwise,
they are anticommutative. In this Majorana fermion representation,
the Hamiltonian (\ref{h}) is mapped to
\begin{eqnarray}
H_{f2}=-J_q\sum_{\langle ij\rangle}W_{ij}-J_p\sum_{\langle
ij\rangle} \tilde W_{ij}+U\sum_{i} Q_i \tilde Q_i,
\end{eqnarray}
where $W_{i,j}=Q_iQ_j$ with $Q_i=i\psi_{i_w}\psi_{i_b}$ and $\tilde
W_{ij}=\tilde Q_i\tilde Q_j$ with $\tilde
Q_i=i\chi_{i_w}\chi_{i_b}$. Since there is no coupling between the
chains, the present model may be transferred to a spin model
\begin{eqnarray} \label{f3}
H_{f3}=-J_q\sum_P  W_P-J_p\sum_P \tilde W_P-U\sum_{i} S_{i_b}^z
S_{i_w}^z,
\end{eqnarray}
by using Jordan-Wigner transformation \cite{jw}:
$\psi_{j_w}=S_{j_w}^y\prod_{j'_s<j_w}S^z_{j'_s}$, 
$\psi_{j_b}=S_{j_b}^x\prod_{j'_s<j_b}S^z_{j'_s}$, 
$\chi_{j_w}=S_{j_w}^x\prod_{j'_s<j_w}S^z_{j'_s}$, and 
$\chi_{j_b}=S_{j_b}^y\prod_{j'_s<j_b}S^z_{j'_s}$,
where $S^{x,y,z}$ are the corresponding Pauli matrices for the
defined Majorana fermions, $W_P=S^y_1 S_2^x S^z_3 S^y_4 S_5^x S^z_6$
and $\tilde{W}_P=S^x_1 S_2^y S^z_3 S^x_4 S_5^y S^z_6$ with the site
labels within a plaquette depicted in Fig. \ref{map1}(d). The order
of the sites is defined as follows: $j_s>j_t$ if the zigzag line
[one of such lines is indicated in Fig. \ref{map1}(b)] including
$j_s$ is higher than that of $j_t$, or if $j_s$ is on the right hand
of $j_t$ when they are in the same line. It is straightforward to
check $[W_P,\tilde W_{P'}]=0$ for every plaquette, and all of them
also commute with the third term. In this spin model, the ground
state can be written as
\begin{eqnarray} \label{g}
|G\rangle=\prod_{P}(1+ W_P)(1+\tilde W_P)|\phi\rangle,
\end{eqnarray}
where $|\phi\rangle=|1\cdots1\rangle$ is a reference state and each
"1" means the eigenvalue of $S^z_{j_{b(w)}}$ being 1. Similar to the
Kitaev's honeycomb model \cite{kitaev06}, all excitations here may
be labelled by two quantum numbers $W_P$ and $\tilde W_P$. The low
energy excitations fall into two closed subsets,  each can be graded
by a $Z_2\times Z_2$ group. The fusion rules of these excitations
are equivalent to the excitations in the toric code model
\cite{kitaev97}, and different graded vortices are mutual abelian
semions \cite{note}.


\begin{figure}[tbp]\centering
\includegraphics[width=5cm]{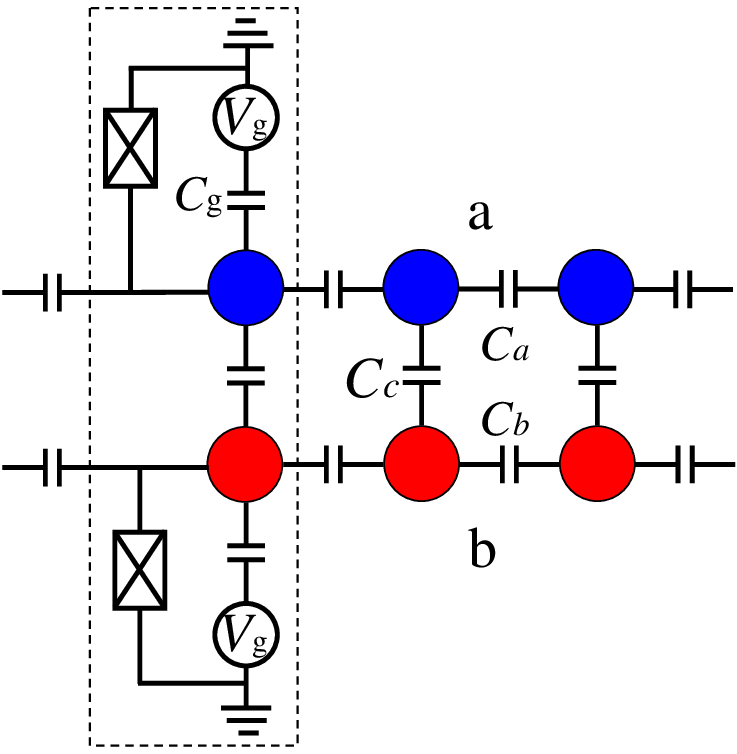}
\caption{(Color online) A schematic circuit of two chains of
capacitively coupled superconducting devices, labeled by $a$ and
$b$, to implement a chain of two-component fermions. Here only  the
first device of the two chains is explicitly shown, while others are
simply denoted as the filled circles with different colors (shades)
for different chains.} \label{cjj} \vspace{-.3cm}
\end{figure}

We now proceed to implement the model with capacitively coupled
superconducting devices, i.e., the Cooper pair box. The key idea is
to use two chains of capacitively coupled superconducting devices,
as shown in Fig. \ref{cjj}, to implement a chain of two-component
fermions. A building block of our implementation, as shown in the
rectangle of Fig. \ref{cjj}, is the two capacitively coupled
superconducting devices. A typical design of a Cooper pair box
consists of a small superconducting island with $n$ excess Cooper
pair charges connected by a Josephson junction with coupling energy
$E_J$ and capacitance $C_J$. A control gate voltage $V_g$ is applied
via a gate capacitor $C_g$.  To quantize the circuit equation, we
first introduce the Hamiltonian and then convert the classical
momentum variable to the momentum operator. Then the Hamiltonian
reads
\begin{eqnarray} \label{h1}
H_{q1}=\sum_{\eta} \left[\frac{\alpha C_{t}^\eta}{2}
(\dot{\varphi}^\eta)^2-E_J^\eta \cos\varphi^\eta \right]-\alpha C
\dot{\varphi}^a\dot{\varphi}^b,
\end{eqnarray}
where $\varphi^\eta$ is the gauge phase drop across the
corresponding junction, $C_t^\eta=C_0^\eta+C_c$ with
$C_0^\eta=C_g^\eta+C_J^\eta$, $\alpha=\left(\hbar /2e\right)^2$, and
the induced charge $n_g^\eta=C_g^\eta V_g^\eta/2e$. At temperatures
much lower than the single-pair charging energy, i.e., $k_B T\ll
E_c^\eta=e^2/(2C_0^\eta)$, and restricting the gate charge to the
range of $n_g \in [0,1]$, only a pair of adjacent charge states
$\{|0\rangle,|1\rangle\}$ on the island are relevant. The
Hamiltonian (\ref{h1}) is then reduced to \cite{youcapacitive}
\begin{eqnarray}\label{2c}
H_{q2}=-{1 \over 2} \sum_{\eta}
\left[\epsilon^\eta\left(1-2n_g^\eta\right)\sigma_z^\eta+
\Delta^\eta \sigma_x^\eta\right]+\lambda\sigma_z^a\sigma_z^b,
\end{eqnarray}
where $\epsilon^{a(b)} =2e^2(C_t^{b(a)}+C_c)/\Lambda$ with
$\Lambda=C_t^a C_t^b-C_c^2$, $\Delta^\eta=E_J^\eta $, $\lambda=e^2
C_c/\Lambda$, $\eta\in\{a, b\}$ and $\sigma_{x,z}$ denotes the
corresponding Pauli matrix in the basis of
$\{|0\rangle,|1\rangle\}$. The single-device terms in Hamiltonian
(\ref{2c}) can be tuned to be zero by conventional methods
\cite{youpt}. Therefore, in what follows, we do not take them into
consideration. For two identical devices ($C_0^\eta=C_0$),
$\epsilon=4E_c$ and $\lambda=e^2 C_c/(C_t^2-C_c^2)\simeq 2\beta E_c$
with $\beta=C_c/C_0$. It is notable that the strength of this
interaction, proportional to the coupling capacitance,  is stronger
than any other present-known coupling methods.

The circuit Hamiltonian of the two coupled chains, as depicetd in
Fig. \ref{cjj}, can be obtained in a similar way. In this extended
multipartite coupling case, the long-range interaction between the
devices would appear, which decays exponentially as $\beta^{|i-j|}$
with $i$ and $j$ being the site labels of the two involved devices
\cite{longrange}. Therefore, the long-range interaction is
negligible if $\beta\ll1$, and in typical experiments
$\beta\simeq0.05$ \cite{tsai}.  Up to the first order of $\beta$,
the interaction Hamiltonian is given by
\begin{eqnarray}\label{2chain}
H_{JJ}=\sum_{\eta;j}\lambda_{\eta}\sigma_{z;j}^\eta\sigma_{z;(j+1)}^\eta
+\sum_{j}\lambda_{c}\sigma_{z;j}^a\sigma_{z;j}^b,
\end{eqnarray}
where $\lambda_{\eta}\simeq e^2C_{\eta}/[C_0+2(C_c+2C_{\eta})]$ and
$\lambda_{c}\simeq e^2C_{c}/[C_0+2(C_c+C_a+C_b)]$.

Once the coupled chains are placed according to the geometry of Fig.
\ref{map1}(a), a corresponding two-dimensional square lattice model
is constructed. Drawing an analogy between the device states in
chain $a$ ($b$) and spin $\uparrow$ ($\downarrow$), the above
interaction Hamiltonian in the addressed system may be rewritten as
the two-component fermion model of Eq. (\ref{h}), with the
parameters $(\lambda_{a}, \lambda_{b}, \lambda_{c})$ corresponding
to $(J_q, J_p, U)$  \cite{note2}. As a result, the topologically
protected ground state of Eq. (\ref{g}) may be implemented with the
present setup of superconducting devices.


In addition, to accomplish certain topological quantum manipulation
tasks, including the examination of the anyon statistics, it is a
must to have a set of basic operations of the devices. In the
Majorana fermion representation, the wanted operations are \cite{yu}
\begin{subequations} \label{m}
\begin{eqnarray} \label{m1}
S^z_{j_b}=i\chi_{j_b}\psi_{j_b}, \quad
S^z_{j_w}=i\chi_{j_w}\psi_{j_w},
\end{eqnarray}
\begin{eqnarray}\label{m2}
S^x_{j_b}=\psi_{j_b}\prod_{j<j_b}(i\chi_j\psi_j), \quad
S^x_{j_w}=\chi_{j_w}\prod_{j<j_w}(i\chi_j\psi_j),
\end{eqnarray}
\end{subequations}
where $S^z_j$ is the spin-flip operator for a given site and it also
transfers a double fermion occupancy to empty or vice versa. $S^x_j$
denotes a nonlocal operation that creates or annihilates a fermion
at site $j$ and  changes the site occupation for sites $j<j_{b(w)}$
\cite{yu}. In the present model, $S^z$ and $S^x$ are effective Pauli
matrices, which, according to Ref. \cite{kitaev06}, may create and
move the excitations.

\begin{figure}[tbp] \centering
\includegraphics[width=4.5cm]{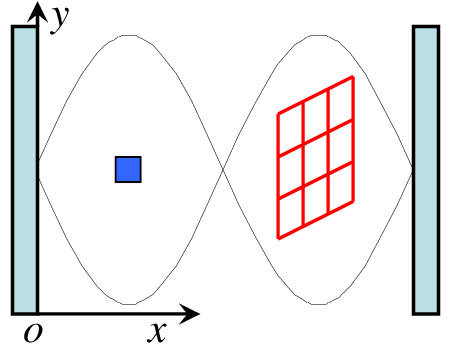}
\caption{(Color online) A schematic diagram of the cavity-assisted
manipulation. The $x$ and $y$ coordinates are denoted by the arrows,
and the $z$ direction is pointing out the $xoy$ plane. The square
lattice (red) is placed to be parallel to the $yoz$ plane. The
square lattice and the auxiliary device (blue rectangle) are placed
with the $x$ coordinate at the antinodes of the single-mode
standing-wave cavity. All superconducting devices are placed with
their loop plane being parallel to the $xoz$ plane, which is
perpendicular to the magnetic component of the cavity field, letting
it be the only contributed component.}  \label{cavity}
\end{figure}

At this stage, let us elaborate how to obtain the wanted operations
in Eqs.~(\ref{m1}) and (\ref{m2}). Notably, individual
addressability is normally a prerequisite in such manipulation.  In
the present proposal, the size of the device setup is macroscopic,
thus individual addressability is taken as granted. To manipulate
the states, we put the lattice into a microwave cavity, with the
geometry of the hybrid system being explained in the caption of Fig.
\ref{cavity}. For simplicity, we consider only the single-mode
standing wave cavity. To be more specific, we reach the cavity
assisted manipulation by a magnetic drive \cite{switch}. The
interaction can be switched on/off by modulating the external
magnetic field to be ac/dc \cite{switch}. With the dc magnetic flux,
the external flux is merely used to address separately the
single-qubit rotations. Under the cavity field,  it is also readily
possible to tune off single-qubit terms. The wanted operations in
Eq. (\ref{m1}) for selected device can be achieved with the cavity
mediated integrations by tuning the driven magnetic flux of the
device  to be of ac. The Pauli matrices $S^z_{jw}$ and $S^z_{jb}$ in
Eq. (\ref{m1}) correspond to $\sigma_x\otimes\sigma_x$ and
$\sigma_y\otimes\sigma_y$ two-body interaction of two devices of
$j$th site, respectively. These two type interactions for each
lattice site can be directly engineered in our hybrid implementation
\cite{switch} as it allows selected addressing of designated
devices. These two interactions are mediated by the virtue cavity
photon, thus we need to keep the cavity mode in the vacuum state.
Here, the devices work in their degeneracy points.

The common cavity mode can also be used to realize the global
stringlike $S^x$ operators in Eq. (\ref{m2}). The off-resonant
interaction between the cavity mode and the selected devices is
\cite{scully}
\begin{equation} \label{qnd}
H_{QND}=\chi n_c  \sum_{j}\sigma_{j}^{z},
\end{equation}
where $n_c=a^{\dag}a$ is the photon number operator of the cavity
mode, the coupling strength is $\chi=g^{2}/2\delta$ with $g$ as the
single-photon Rabi frequency for the cavity mode, and $\delta$ is
the detuning between the cavity mode frequency $\omega_c$ and
optical transition frequency in atomic spins. In our implementation,
this can be the $|1\rangle\rightarrow|2\rangle$ transition of the
selected devices, where $|2\rangle$ is an ancillary energy level
beyond the qubit subspace $\{|0\rangle, |1\rangle\}$, and the
frequency of the drive ac flux satisfies
$\omega=\widetilde{\omega}_{12}+\omega_c+\delta$ with
$\widetilde{\omega}_{12}=2E_c(3-2n_g)/\hbar$.  To avoid the
transitions $|0\rangle\rightarrow|1\rangle$ and
$|0\rangle\rightarrow|2\rangle$ by the ac drive, we tune
$\widetilde{\omega}_{01}=2E_c(1-2n_g)/\hbar$ via $n_g$ so that
$\delta \ll \Delta_{1,2}$, where
$\Delta_1=\omega-\widetilde{\omega}_{01}$ and
$\Delta_{2}=\widetilde{\omega}_{01}+\widetilde{\omega}_{12}-\omega$
are the corresponding detunings. This quantum non-demolition (QND)
Hamiltonian (\ref{qnd}) preserves the photon number of the cavity
mode. Within the $n_c\in\{0, 1\}$ subspace, the evolution of the QND
Hamiltonian during the interaction time $\tau=\pi/ 2\chi$ yields
\cite{jiang}
\begin{align} \label{cs}
U  =\exp\left[ -iH\tau\right] =\left\{
\begin{tabular}{cc}
$\mathbf{I}$ & ~~~for $n_{c}=0$ \\
$\left( -i\right)^{N}\prod_{j}\sigma_{j}^{z}$ & ~~~for $n_{c}=1$%
\end{tabular} \right.
\end{align}
where $N$ is the number of the selected devices. From Eq.
(\ref{cs}), (controlled) string operations for an arbitrary string
can be achieved \cite{jiang}.

If the cavity is initially prepared in the $n_c=1$ state, the global
operation reduces to the string operation
$U_z=\prod_{j}\sigma_{j}^{z}$. As all string operators are
equivalent to $U_z$ up to local single spin rotations \cite{jiang},
all string operations for arbitrary string can be achieved:
$U_x=\prod_{j}\sigma_{j}^{x}=HU_zH$ and
$U_y=\prod_{j}\sigma_{j}^{y}=RU_zR$, where $H=\prod_{j}H_{j}$ and
$R=\prod_{j}R_{j}$ with
$H_j=\left(\sigma_j^x+\sigma_j^z\right)/\sqrt{2}$ being the Hadamard
rotation and $R_j=\exp\left(-i{\pi \over 4} \sigma^z_j\right)$.
Therefore, with this elementary operation, creation and manipulation
of anyons are likely feasible in our scheme. For example,
\begin{eqnarray}
S^x_{j_b}&=&\psi_{j_b}\prod_{k_b<j_b}(i\chi_{k_b}\psi_{k_b})
\prod_{k_w<j_b}(i\chi_{k_w}\psi_{k_w})\notag\\
&=&\sigma_{y;j}^b\prod_{k_b<j_b}(i\sigma_{x;k_b}^a\sigma_{x;k_b}^b)
\prod_{k_w<j_b}(i\sigma_{y;k_w}^a\sigma_{y;k_w}^b),
\end{eqnarray}
denotes a nonlocal operation that could create a domain-wall-like
excitation/semion (at the site $j$) from the ground states under the
open boundary condition~\cite{yu}.

If the cavity is initially prepared in a superposition of zero- and
one-photon states, the global operation in Eq. (\ref{cs}) reduces to
a controlled-string operation: $U_{cs}=\mu |0\rangle\langle0|\otimes
I+ \nu|1\rangle\langle1|\otimes U_z$, where the parameters $\mu$ and
$\nu$ are controlled by the initially prepared photon number state.
With such a controlled-string operation, one is able to access the
topological memory and to build anyonic interferometry \cite{jiang}.

In simulating the string operations, we need to engineer the cavity
number states. Therefore, beside the square lattice, we also place
an ancilla device in the cavity, as shown in Fig. \ref{cavity},
which is used to control the cavity photon number state by swapping
its states with that of the cavity using  the resonate cavity-device
interaction. This swap operation can be achieved by the famous
Jaynes-Cummings model Hamiltonian: $H_{JC}=\Omega\left(a\sigma^+
+a^{\dag}\sigma^-\right)$, which can be implemented in our system by
choosing the frequency of the ac driven magnetic flux for the
ancillary device satisfying
$\omega=\omega_c+\widetilde{\omega}_{01}$. In this case, we need to
tune the device slightly away from the degeneracy point, which
results in a shorter decohenrence time. Fortunately, the resonate
operation is also much faster.


In summary, we have proposed an exotic scheme to implement a
two-component fermion model using superconducting quantum circuits,
which was shown to support abelian anyonic excitations. Most
intriguingly, we have elaborated how to achieve all the wanted
operations that could create and manipulate the anyonic states. Our
approach provides an experimentally feasible scenario to access the
topological memory and to build the anyonic interferometry.


We thank Yue Yu, L. B. Shao  and Liang Jiang for many helpful
discussions. This work was supported by the RGC of Hong Kong under
Grants No. HKU7045/05P and No. HKU7049/07P, the URC fund of HKU, the
NSFC under Grants No. 10429401, No. 10674049 and No.  10625416, and
the National Basic Research Program of China (No. 2006CB921800, No.
2007CB925204  and No. 2009CB929300).


\begin{thebibliography}{99}
\bibitem{note} Although it is hard to figure out the presence of anyons directly from
Hamiltonian (\ref{h}), it is quite clear from the theory of Kitaev
\cite{kitaev06} that Hamiltonian (\ref{f3}) does support anyonic
excitations from the ground state denoted by Eq. (\ref{g}).
\bibitem{note2} It is noted that the signs before $\lambda_a$ and $\lambda_b$ in Hamiltonian (7)
are intially positive. However, the signs  can be tuned to be minus
by a proper redefinition of the analogy between device states
$\{|0\rangle, |1\rangle\}$ and pseudospin states $\{|+\rangle,
|-\rangle\}$, i.e., $|0\rangle$ ($|1\rangle$) corresponds to
$|+\rangle$ ($|-\rangle$) for all odd-number sites while $|0\rangle$
($|1\rangle$) corresponds to $|-\rangle$ ($|+\rangle$)  for all
even-number sites.
\end{thebibliography}
\end{document}